\documentclass[preprint,showpacs,prb,amsmath]{revtex4}

\usepackage{graphicx}
\usepackage{bm}
\def\bsigma{\bm{\sigma}}

\begin{document}
\title{Dynamics of Dissipative Quantum Hall Edges}
\author{M.D. Johnson}
\email{mjohnson@ucf.edu}
\affiliation{Department of Physics,
University of Central Florida, Orlando, FL 32816-2385, USA}
\author{G. Vignale}
\affiliation{Department of Physics, University of Missouri, Columbia, MO 65211, USA}
\date{March 4, 2003}
\begin{abstract}
We examine the influence of the edge electronic density profile and
of dissipation on edge magnetoplasmons in the quantum Hall regime,
in a semiclassical calculation.
The equilibrium electron density on the edge, obtained using a
Thomas-Fermi approach, has incompressible stripes produced by energy
gaps responsible for the quantum Hall effect. We find that these stripes have
an unobservably small effect on the edge magnetoplasmons. But
dissipation, included phenomenologically in the local conductivity,
proves to produce significant oscillations in the strength and
speed of edge magnetoplasmons in the quantum Hall regime.
\end{abstract}
\maketitle

\section{Introduction}
\label{sec:intro}
We investigate electron dynamics on the edges of a two-dimensional
electron system in the quantum Hall\cite{qhe} regime. Our aim is to examine 
the influence of the edge electronic density profile and of dissipation on the
low-energy dynamics. Dealing with the combined effects of electron interactions,
disorder, and density inhomogeneities is very complicated. Consequently
we adopt a simple semiclassical edge magnetoplasmon model, 
appropriate for smooth edges, with an
eye towards explaining the kinds of phenomena that can be expected. 
This work helps explain some recent dynamical measurements, and may
help develop such measurements into a quantitative tool to examine
the structure and properties of quantum Hall edges.

This work was motivated by recent time-of-flight experiments by Ernst
{\itshape et al.}\cite{e1,e2,e3} on quantum Hall bars. These consist
of rectangular two-dimensional (2D) electron systems with strong
transverse magnetic fields $B$. In these
experiments a voltage pulse sent down a thin ($\sim 1\,\mu$m) voltage
probe triggers a density pulse. This pulse moves along the edge
and is then detected as a voltage signal at a second probe. The
density pulse's transit time was found to generally increase with
magnetic field (a behavior expected classically), but superimposed on
this trend were sudden changes associated with the quantum Hall 
effect\cite{e2,e3}.
The strength of the detected signal showed corresponding variations.
A similar technique was used earlier by Ashoori {\itshape et al.} to
look for counter-propagating modes on a single mesa\cite{ashoori,mjz}.

The appropriate framework for studying electronic properties on the
edges of Hall bars depends on the ratio of two length scales.
These are the edge width $a$ (the distance over which the
electron density falls from its bulk value to zero) and
the magnetic length $\ell=\sqrt{\hbar c/eB}$ (of order 10 nm
for typical magnetic fields).
The limiting cases are of {\itshape sharp}
edges ($a\sim \ell$) and {\itshape smooth} edges ($a\gg\ell$).
Sharp edges can can be obtained by cleaving\cite{chang}.
The electronic properties of sharp edges
are largely quantum mechanical in nature, and determined
by the state of the bulk. For example, at filling factors $\nu$ 
($=2\pi\ell^2$ times the bulk two-dimensional density)
at which the fractional quantum Hall effect (FQHE) is exhibited,
the edge dynamics may be describable in terms of chiral one-dimensional
Luttinger liquids\cite{wen,stone,kane,pal,shytov}.
This possibility has received
some experimental support (with a few surprising twists) in measurements of
tunneling onto 
edges\cite{chang,milliken,grayson,conti,zul,lee,levitov,grayson2}.

In this paper we deal only with the opposite limit of smooth edges,
$a\gg\ell$. This is typical in samples which are not cleaved.
In the absence of cleaving, the edge structure is determined by the 
electrostatics of the 2D electron system, donors, and gates.
These produce edges with a typical width $a$ of order $1\,\mu$m.
This is much greater than the magnetic length $\ell$, so 
electrostatically-defined edges are smooth. Since the electrostatics
dominates, the electron density on the edge is close to 
the classical solution: a compressible electron gas with a density
falling smoothly from its bulk value to zero over a distance $a$.
Quantum mechanical energy gaps arising in the quantum Hall
regime modify this picture, producing incompressible stripes along the
edge at certain filling factors\cite{beenakker,chk}.
Under these circumstances the low-energy edge dynamics is determined largely
by the compressible portion of the edge electron density.
Then a reasonable framework for
beginning a study of the edge dynamics is the semiclassical acoustic edge
magnetoplasmon\cite{VM,AG}. This is the direction we follow.

In the next section we will summarize existing work which leads to
the edge magnetoplasmon. The new results of this paper are then
contained in Sections~\ref{sec:profile} and \ref{sec:dissipation}.
In Section~\ref{sec:profile}
we examine the effect of incompressible stripes on the edge magnetoplasmon.
The variation of stripe position and width with magnetic field modifies
the edge magnetoplasmon's velocity, but the effect appears too small to
be observable. In Section~\ref{sec:dissipation} we examine how 
disorder-induced dissipation both damps and slows the edge modes. 
In the quantum Hall regime dissipation varies with magnetic field,
and so the edge magnetoplasmon's strength and speed should change
simultaneously. We find that this effect is observably large.

\section{Edge Magnetoplasmons}
\label{sec:emp}
Edge magnetoplasmons are classical or semiclassical electronic density
fluctuations on the edge of a 2D electron system, in the presence of a
magnetic field. They are acoustic modes, with frequencies much less than
the classical cyclotron frequency $\omega_c=eB/mc$.
Edge magnetoplasmons have been studied thoroughly in two cases.
Volkhov and Mikhailov used a Wiener-Hopf technique to investigate their
properties for half spaces (the extreme sharp-edge limit, $a\rightarrow0$),
in the presence of dissipation\cite{VM}. Later Aleiner and Glazman
studied smooth edges in the absence of dissipation, in strong magnetic
fields\cite{AG}. They found that the smooth edge can support many modes,
with differing numbers of nodes; the fastest, nodeless mode corresponded
to the Volkhov-Mikhailov edge magnetoplasmon. In this paper we extend these
studies to investigate the roles of differing edge density structures
and dissipation in the case of smooth edges. 
We begin with a summary of the basics of edge magnetoplasmons, following
Aleiner and Glazman\cite{AG}.

\begin{figure}
\begin{center}
\includegraphics*[scale=0.5]{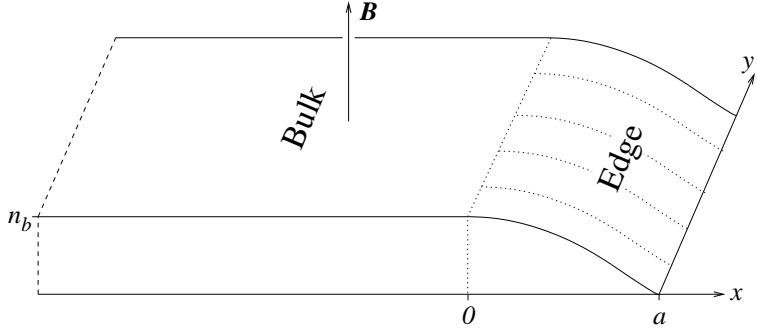}
\end{center}
\caption{Schematic equilibrium electronic density $n_0(x)$ near the
edge of a Hall bar. The system is taken to be semi-infinite in the transverse
($x$) direction and infinite in the longitudinal ($y$) direction. 
In the bulk ($-\infty<x<0$) the density takes the constant value $n_b$.
On the edge $(0\le x\le a$) the density falls from the bulk
value to zero.}
\label{geometry}
\end{figure}

Consider a Hall bar, a 2D electron system in the $xy$ plane 
subject to a perpendicular magnetic field $\mathbf{B}$ (Fig.~\ref{geometry}).
To study a single edge we can
take the system to be infinite along the edge (coordinate $y$) and
semi-infinite in the transverse direction (coordinate $x$).
In equilibrium this system is translationally invariant along the edge,
and so the equilibrium 2D electronic density $n_0(x)$ depends only on
the transverse coordinate $x$.
We can distinguish {\itshape bulk} and {\itshape edge} regions.
In the bulk $n_0(x)$ takes a constant bulk value $n_b$, and on
the edge $n_0(x)$ falls from $n_b$ to zero. We take the bulk and
edge regions to be, respectively, $-\infty < x < 0$ and $0<x<a$.

We also allow the presence of a top gate (a grounded piece of metal
parallel to the 2D electron system, a distance $d$ away). The only
mobile charges in the system are the electrons in the 2D plane and
in the gate (the latter included as image charges). 
Classically the equilibrium density
$n_0(x)$ corresponds to a vanishing in-plane electric field.
If the electronic density takes a nonequilibrium value
$n({\mathbf{r}})=n_0(x)+\delta n({\mathbf{r}})$, then the net electric
field in the 2D plane, including image charges, becomes
\begin{eqnarray}
\mathbf{E}(\mathbf{r}) &=& - \nabla \Phi(\mathbf{r}),\nonumber\\
\Phi(\mathbf{r}) &=& -\frac{e}{\varepsilon} \int d^2 r' \,
\delta n( {\mathbf{r}'} ) \left( \frac{1}{|\mathbf{r}-\mathbf{r}'|}
- \frac{1}{[ {|\mathbf{r}-\mathbf{r}'|^2} + (2d)^2 ]^\frac{1}{2}}
\right).
\label{E}
\end{eqnarray}
Here $-e$ is the electronic charge, and $\varepsilon$ is the dielectric
constant of the material within which the 2D electron system lies
(12.4 for GaAs).
This electric field expression is also satisfactory for time-dependent
densities $\delta n(\mathbf{r},t)$, provided they vary sufficiently slowly.

This electric field produces a restoring force which leads to oscillations,
{\itshape i.e.}, magnetoplasmons. Two more ingredients are needed to see this.
Classically the electric field induces a current density $\mathbf{j}$
via a local conductivity tensor $\bsigma$:
\begin{equation}
\label{j}
\mathbf{j}(\mathbf{r}) = \bsigma(\mathbf{r}) \cdot \mathbf{E}(\mathbf{r}).
\end{equation}
Dynamics enters via the continuity equation,
\begin{equation}
\label{continuity}
\nabla \cdot {\mathbf{j}} + \frac{\partial (-e \delta n)}{\partial t} = 0.
\end{equation}
Inserting Eqs.~(\ref{E}) and (\ref{j}) into Eq.~(\ref{continuity}) gives
an equation of motion for $\delta n({\mathbf{r}},t)$. One seeks
wave solutions along the edge by writing
\begin{equation}
n({\mathbf{r}},t) = n_0(x) + \delta n({\mathbf{r}},t) =
n_0(x) + f(x) e^{i(ky-\omega t)}.
\end{equation}
Linearizing in the edge mode density $f(x)$ then produces an
integrodifferential equation:
\begin{subequations}
\label{modeeqn}
\begin{eqnarray}
e\omega f(x) &=& \left\{ k\frac{\partial\sigma_{xy}}{\partial x}
-i \left[ \sigma_{xx} \frac{\partial^2 }{\partial x^2} + 
\frac{\partial\sigma_{xx}}{\partial x}
\frac{\partial}{\partial x}-k^2\sigma_{xx}\right]\right\}\phi(x),\label{feqn}\\
\phi(x) &=& -
\frac{e}{\varepsilon}\int_{-\infty}^{a} dx'\,f(x')\left[
K_0(|k(x-x')|) - K_0(|k|\sqrt{(x-x')^2+(2d)^2})\right].
\label{phieqn}
\end{eqnarray}
\end{subequations}
Here the Bessel function $K_0$ results from
performing the integral over $y'$ in Eq.~(\ref{E}). 
Eqs.~(\ref{modeeqn}) are to be solved for the mode densities
$f(x)$ and corresponding eigenfrequencies $\omega$.
In general for fixed $k$ there are many solutions, \textit{i.e.} modes
with different numbers of nodes $n=0,1,2,\dots$\cite{AG}. Solving
Eqs.~(\ref{modeeqn}) with
varying $k$ produces each mode's dispersion relation $\omega_n(k)$. This
then yields the mode's speed and signal strength as a function of
magnetic field.

The conductivity tensor components
$\sigma_{xx}, \sigma_{xy}$ appearing in the linearized equation
(\ref{feqn}) are those for the system at its equilibrium density
$n_0(x)$. Semiclassically these are found from the electron's equation
of motion:
\begin{equation}
{\mathbf{F}} = -e \left({\mathbf{E}} + \frac{1}{c}{\mathbf{v}}\times{\mathbf{B}}
\right) - \frac{m}{\tau}{\mathbf{v}} = m\dot{\mathbf{v}},
\end{equation}
where $m$ is the effective mass ($0.067$ m$_e$ in GaAs).
Consider a time dependence 
${\mathbf{v}}(t)={\mathbf{v}}e^{-i\omega t}$ and a
constant magnetic field $\mathbf{B}$ in the $-z$ direction. Solving
for the 2D velocity components $v_x,v_y$
and writing the linearized current density as
${\mathbf{j}}({\mathbf{r}})=-e n_0(x) {\mathbf{v}}$ gives the classical
Drude result:
\begin{eqnarray}
\sigma_{xy}(x,\omega) &=& \frac{e^2n_0(x)}{m}\, 
\frac{\omega_c}{\omega_c^2 - (\omega + i/\tau)^2},\nonumber\\
\sigma_{xx}(x,\omega) &=& \frac{e^2n_0(x)}{m}\, 
\frac{-i\omega+1/\tau}{\omega_c^2 - (\omega + i/\tau)^2}.
\label{sigma}
\end{eqnarray}
Here $\omega_c=eB/mc$ is the classical cyclotron frequency.
Dissipation enters via the relaxation time $\tau$.

An important assumption we make in this paper is that the relaxation time
$\tau$ does not depend on position, \textit{i.e.}, it is a {\it global property}
of the edge, determined only by the bulk density of the fluid. The physical
idea is that momentum relaxation along the edge is a nonlocal process,
controlled by long-range interactions with low-energy excitations both
in the bulk and in the edge. Since the density of low-energy bulk
excitations is much lower when the bulk is in an incompressible state
than when it is not, the momentum relaxation rate $1/\tau$ is
expected to exhibit periodic oscillations, becoming very small on the
quantum Hall plateaus, and growing to sizeable values in the transitions
between the plateaus. 
An order of magnitude estimate of $1/\tau$ is given by
the observation that the DC longitudinal and transverse conductivities
become comparable in magnitude between plateaus,
$\sigma_{xx} \sim \sigma_{xy}$. By Eq.~\ref{sigma} this yields
$1/\tau \sim \omega_c$. In the quantum Hall regime
$2\pi\ell^2n_b\sim1$. Consequently the estimate $\omega_c\tau\sim1$
also gives a Drude conductivity equal to the typical
longitudinal conductivity between plateaus:
$\sigma_{xx} = n_be^2\tau/m \sim e^2/h$.

The mode equation (\ref{modeeqn}) has solutions which live on the edge.
This is easiest to see in the case of no dissipation, $1/\tau=0$.
In the strong-magnetic-field limit, solutions of Eqs.~(\ref{modeeqn})
then exist with $\omega\ll\omega_c$.
In this case $\sigma_{xx}$ can be neglected and there is no dissipation.
(This is the case examined by Aleiner and Glazman to study edge
magnetoplasmons in the quantum Hall regime\cite{AG},
modified slightly here to include a top gate.)
From Eqs.~(\ref{modeeqn},\ref{sigma}) one sees that when
$\sigma_{xx}=0$ the mode density $f(x)$ is
nonzero only on the edge ($0<x<a$), where the equilibrium density
is nonuniform. For each $k$ there are many modes, with differing
numbers of nodes and wave velocities\cite{AG}. We will
concentrate on the fastest mode, the one which triggers the 
signal onset in time-of-flight experiments. 

In the absence of dissipation, the fastest mode is always nodeless.
The following is a simple estimate of the speed when $d\ll a$.
For $ka\ll1$, the term in square brackets in Eq.~(\ref{phieqn})
is approximately $\log[1+(2d/(x-x'))^2]/2$. For $d\ll a$ this is
strongly peaked at $x'=x$, and so 
\begin{equation}
\phi(x) \approx -\frac{e}{2\varepsilon}f(x)\int_{-\infty}^{\infty} 
\log\left[ 1 + \left(\frac{2d}{x-x'} \right)^2 \right] = 
-\frac{4\pi e d}{\varepsilon} f(x) .
\label{lca}
\end{equation}
This is known as the local capacitance approximation\cite{zhitenev}. 
Inserting this into Eq.~(\ref{feqn}), and putting $d\sigma_{xy}/dx\sim
\sigma_{xy}^{bulk}/a$ then gives the estimate\cite{zhitenev}
\begin{equation}
\omega = sk, \text{ with } s\approx \frac{4\pi e^2 n_b}{\varepsilon m \omega_c}
\frac{d}{a}.
\label{sest}
\end{equation}
Thus classically, in the absence of dissipation, for gates very close
to the 2D electron system, the edge magnetoplasmon speed $s$ varies as
$d/a$, and moreover decreases with increasing magnetic field.
The top gate is useful experimentally because it slows the edge
magnetoplasmons\cite{zhitenev}.

In this paper we will examine the effects of dissipation and of
incompressible stripes on the edge modes. The modes are calculated
classically, using Eqs.~(\ref{modeeqn}) with the Drude conductivity
Eq.~(\ref{sigma}). Quantum mechanics enters by influencing the
conductivity tensor in two ways: by adding incompressible stripes to
the edge density $n_0(x)$ and by influencing the phenomenological
relaxation time $\tau$. This combination gives 
a semiclassical calculation of edge magnetoplasmons.

\section{Incompressible stripes on the edge}
\label{sec:profile}

In this section we examine how incompressible stripes on the edge
influence the edge magnetoplasmons. The biggest energy
scale in determining the edge density profile is given by electrostatics,
and consequently the edge density is nearly classical. The structure is modified
somewhat by quantum mechanical energy gaps---kinetic energy gaps due
to the quantization of the kinetic energy into Landau levels, and
interaction energy gaps responsible for the fractional quantum Hall effect.
These energy gaps produce incompressible stripes\cite{beenakker,chk}.

In the case of smooth edges ($\ell \ll a$) a good approximation to
the edge density profile can be obtained using Thomas-Fermi theory,
the simplest form of local density approximation.
Ordinarily the Thomas-Fermi approximation amounts to the statement
that the local electrochemical
potential is a constant value $\mu$ in equilibrium. This must be
modified in a system with incompressible regions. One must find the
density for which
\begin{equation}
T(x) + V_H(x) + V_{xc}(x) + W(x) \le \mu,
\label{tf}
\end{equation}
where the equality holds in compressible regions and the
inequality in incompressible regions.
The leading term $T(x)$ above represents the kinetic energy required to
add one electron, in a local density approximation. 
This is determined by the local filling factor
(the number of Landau levels locally occupied), $\nu(x)=2\pi\ell^2n_0(x)$.
An added
electron would go into the topmost level, and so (including spin)
\begin{equation}
T(x) = \hbar\omega_c\left(\left[ \frac{1}{2}\nu(x) \right] + \frac{1}{2}\right),
\label{t}
\end{equation}
where $[\dots]$ denotes the integer part. 
The second term in Eq.~(\ref{tf}) includes electron-electron interactions in the
Hartree approximation:
\begin{equation}
V_H(x) = \frac{e^2}{\varepsilon} \int d^2 r' \,
n_0( x' ) \left( \frac{1}{|\mathbf{r}-\mathbf{r}'|}
- \frac{1}{[ {|\mathbf{r}-\mathbf{r}'|^2} + (2d)^2 ]^\frac{1}{2}}
\right)\approx U n_0(x),
\label{vh}
\end{equation}
where $U={4\pi e^2 d / \varepsilon}$.
This last very useful simplification is the local capacitance approximation
used in Eq.~(\ref{lca}); it is valid when
the top gate spacing is small ($d/a \lesssim 0.1$).
The term $V_{xc}$ in
Eq.~(\ref{tf}) is the exchange-correlation energy in a local density
approximation. Its inclusion allows us to investigate edges in the
fractional quantum Hall regime. We use a form for $V_{xc}$ taken
from Ref.~\onlinecite{LJH}. The last term $W(x)$ represents an
external confining potential. We assume this to be nonzero only
along the edge ($0<x<a$).
The constant $\mu$ is adjusted to produce the desired total electron number.

\begin{figure}
\begin{center}
\includegraphics*[scale=0.4]{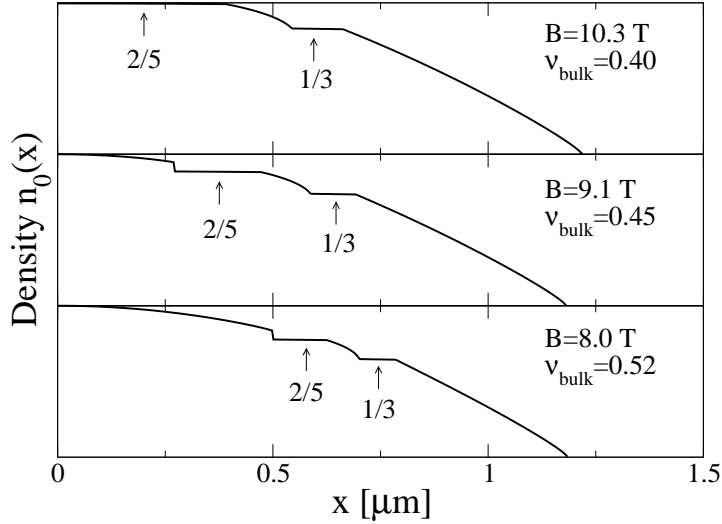}
\end{center}
\caption{Equilibrium electronic densities $n_0(x)$ along a $1.2\,\mu$m edge
for three values of magnetic field. The density falls from its bulk value 
$n_b=10^{11}\,\mathrm{cm}^{-2}$ to zero in each plot. Pictured are two
incompressible stripes at $\nu=1/3$ and $\nu=2/5$. As the magnetic field
increases, the bulk filling factor decreases and the incompressible stripes
move move closer to the bulk. Here $d=30$ nm.
}
\label{densities}
\end{figure}

The solution of Eq.~(\ref{tf}) consists of a sequence of alternating
compressible and incompressible stripes, as pictured in Fig.~\ref{densities}.
It is easy to see how this works in the local capacitance approximation.
For the time being we neglect $V_{xc}$ and therefore consider only the 
spin-unresolved integer quantum Hall effect. Throughout a compressible region
$T(x)$ takes a constant value, and Eqs.~(\ref{tf},\ref{vh}) give a density
\begin{equation}
n_0(x) = (\mu - W(x) - T)/U.
\label{n0}
\end{equation}
Now imagine moving inward through a compressible region in which
$k-1$ Landau levels are filled, and the $k^{th}$ is being filled. 
Then $T=\hbar\omega_c(k+1/2)$. As $x$ moves inward, the confining
potential $W(x)$ decreases, and so the density $n_0(x)$ in Eq.~(\ref{n0})
increases. However at some point $x_o$ the $k^{th}$ Landau level fills
completely. Increasing the density further would require a discontinuous
jump in kinetic energy, and locally the inequality in Eq.~(\ref{tf}) would
be violated. Thus an incompressible stripe forms with local filling
factor $\nu(x)=2k$. The stripe's outer boundary is the value $x_o$ where
the compressible region's local filling factor grows to $2k$, and so
$x_o$ is found from:
\begin{equation}
\hbar\omega_c(k+1/2)+ U 4k\pi\ell^2 + W(x_o) = \mu.
\label{x0}
\end{equation}
The incompressible stripe lasts until $W(x)$ becomes small enough that
electrons can occupy the $(k+1)^{st}$ Landau level. Thus the inner boundary
$x_i$ of the incompressible stripe with $\nu=2k$ is determined by
\begin{equation}
\hbar\omega_c(k+3/2)+ U 4k\pi\ell^2 + W(x_i) = \mu.
\label{xsubi}
\end{equation}
If in the incompressible stripe the slope $W'$ is approximately
constant, then Eqs.~(\ref{x0},\ref{xsubi}) give a stripe width
\begin{equation}
\Delta x = x_o - x_i \approx \frac{\hbar\omega_c}{W'}.
\label{width}
\end{equation}
Although the confining potential $W$ is unknown, we can estimate the stripe width
as follows. The electron density falls from its bulk value $n_b$ to zero
as $x$ goes from $0$ to $a$. Choose a magnetic field so that the bulk
filling factor is less than two. Then $T(x)$ is a constant, and
applying Eq.~(\ref{tf}) at $x=0,a$ gives $Un_b + W(0)=W(a)$. Consequently
\begin{equation}
W' \sim \frac{W(a)-W(0)}{a} \sim \frac{Un_b}{a},
\end{equation}
and, inserting this into Eq.~(\ref{width}),
\begin{equation}
\Delta x \sim \frac{\hbar \omega_c a }{ U n_b}.
\end{equation}
Typical experimental values ($d=130$ nm, $n_b=10^{11}$ cm$^{-2}$) 
give $Un_b = 190$ meV. This is much greater than
the cyclotron energy $\hbar\omega_c$ at accessible fields, and so
the incompressible stripes make up a small fraction of the total edge. 
Eliminating the local capacitance approximation 
changes the picture only slightly: the electrostatic potential $V_H(x)$
varies a small amount across the incompressible stripe, whose width
consequently varies slightly from the above estimate.

Incompressible stripes also arise from interaction gaps,
represented here by $V_{xc}$, the contribution of exchange and correlation
in a local density approximation. The corresponding energy gaps are typically
of order of a fraction of $e^2/\ell$, which is smaller than the cyclotron energy. Consequently the resulting
incompressible stripes are even narrower than resulting from the kinetic
energy quantization. 

Given a particular confinement potential $W(x)$, we numerically
solve Eq.~(\ref{tf}) for the equilibrium density $n_0(x)$. An example
using a parabolic edge confinement is shown in Fig.~\ref{densities}. 
Incompressible stripes of constant density are visible at 
FQHE fractions $\nu=2/5$ and $1/3$. The stripes vary with magnetic field:
as the magnetic field increases the incompressible stripes move inward
(toward the bulk), and new stripes appear at the outer edge.
(To make the incompressible stripes more evident, we
have used a smaller gate spacing $d=30$ nm in Fig.~\ref{densities}).

\begin{figure}
\begin{center}
\includegraphics*[scale=0.35]{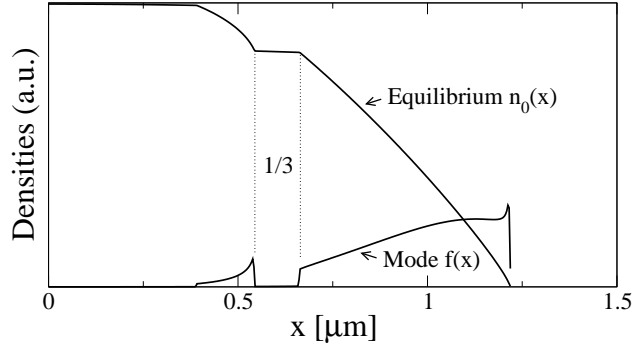}
\end{center}
\caption{The equilibrium density $n_0(x)$ and the edge magnetoplasmon
density $f(x)$ for the fastest mode, with $B=10.3\,\mathrm{T}$ 
($\nu_{\mathrm{bulk}}=0.40$). The edge mode has a nonvanishing density
only where the equilibrium density is compressible. Consequently $f(x)$
drops to zero along the incompressible $\nu=1/3$ stripe. The 
long-range Coulomb force couples all compressible regions.
}
\label{density-and-mode}
\end{figure}

Now we can investigate the incompressible stripes' effect on
the edge magnetoplasmons. 
At each magnetic field we find $n_0(x)$ as outlined above, and then
solve Eqs.~(\ref{modeeqn}) numerically for several values of $k$.
Since our focus in this section is on the role of electronic structure
we neglect dissipation (setting $\sigma_{xx}=0$ in Eq.~(\ref{feqn})); 
numerical solution of Eq.~(\ref{modeeqn}) is then
straightforward. The results are similar to that of
Ref.~\onlinecite{AG}, modified
by the top gate and also by the different form of the equilibrium density.
For each $k$ there are many modes. The fastest
mode is always the nodeless mode. Since time-of-flight experiments
are most sensitive to the fastest mode, we concentrate on it,
finding a set of $k$-dependent mode densities $f(x)$ and
eigenfrequencies $\omega$. An example is shown in Fig.~\ref{density-and-mode}.
Notice that the mode density $f$ extends throughout the entire edge,
dropping to zero at the position of an incompressible stripe but filling
all of the compressible region. This is a consequence of the finite range
of the interaction~(\ref{phieqn}), which reaches across the incompressible stripe.

\begin{figure}
\begin{center}
\includegraphics*[scale=0.35]{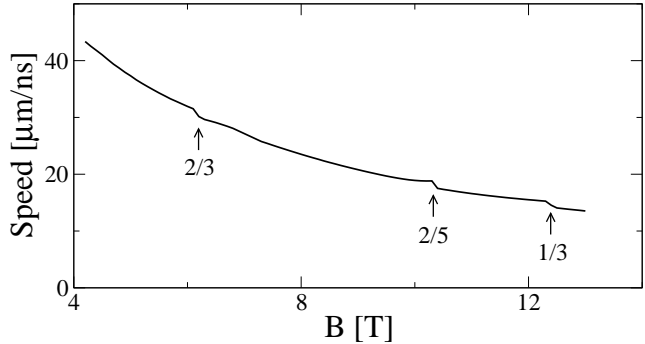}
\end{center}
\caption{Speed of the fastest edge magnetoplasmon for the case shown in
Figs.~\ref{densities} and \ref{density-and-mode}. The merging of
incompressible stripes into the bulk produces small drops in speed.}
\label{speeds-strip}
\end{figure}

In the long-wavelength limit the eigenfrequency 
$\omega\approx sk$ with $s$ the magnetoplasmon speed.
By repeating the calculation with varying $B$ we find
the mode speed as a function of magnetic field.
Results in the FQHE regime are shown in Fig.~\ref{speeds-strip}. 
The speed falls generally as $1/B$, consistent with the classical
result. Superimposed are small variations due to the spatial variation
of the stripes with changing magnetic field. (As $\nu_b$ moves from
just below to just above a FQHE fraction, a relatively large incompressible
stripe shrinks abruptly into the bulk. This enlarges the
compressible region and so diminishes the mode's speed.)

It is evident in Fig.~\ref{speeds-strip} that the $B$-dependence of
the incompressible stripes has only a small influence on the edge
magnetoplasmon's speed. [In fact this is exaggerated in Fig.~\ref{speeds-strip};
for the purpose of illustration we (inconsistently) used $d=130$ nm in Eqs.~\ref{modeeqn}
even though the equilibrium density $n_0(x)$ was calculated with $d=30$ nm. This gives
a stronger restoring force and a larger variation in mode speed.]
Within this semiclassical model, incompressible stripes do not lead
to observable variations in speed.

\section{Dissipation}
\label{sec:dissipation}

In this section we examine how dissipation affects edge magnetoplasmons.
We find that variations in dissipation due to the physics of the
quantum Hall effect produce simultaneous variations in the edge
mode's speed and strength which should be observably large.

In the previous section we showed that incompressible stripes do not
significantly influence the edge magnetoplasmons. Consequently in this
section we use smooth edge density profiles, neglecting incompressible stripes.
This lets us focus on the role of dissipation.

For orientation we begin with a simplified model which is analytically
solvable, and which exhibits the effect of dissipation in a transparent
way. This helps us disentangle the results of a numerical solution of the
mode equations (\ref{modeeqn}), which we will present last.

\subsection{Linear model}
\label{subsec:model}

The model in question consists of: (i) taking the equilibrium edge
density to be linear, and (ii) calculating the electric potential $\phi$
in the local capacitance approximation [Eq.~(\ref{lca})].

Write the equilibrium density as $n_0(x) = n_b \overline{n}(x)$ with 
\begin{equation}
\overline{n}(x) = \left\{ \begin{array}{ll}1 & x\le 0\\
1-x/a\quad& 0\le x\le a.\end{array}\right.
\end{equation}
Using this density profile, the local capacitance approximation Eq.~(\ref{lca}),
and the conductivities in Eq.~(\ref{sigma}),
the mode equation [Eq.~(\ref{modeeqn})] becomes
\begin{eqnarray}
\lefteqn{\frac{\omega}{\omega_c}
\left[ 1-\left(\frac{\omega}{\omega_c}+\frac{i}{\omega_c\tau}\right)^2\right]
f(x) =} \nonumber\\
& & \gamma a^2 \left[ -k \overline{n}' f +
         \left(\frac{\omega}{\omega_c}+\frac{i}{\omega_c\tau}\right)
\left( \overline{n}f'' + \overline{n}'f' - k^2\overline{n}f
\right) \right].
\label{modede}
\end{eqnarray}
This is written to make the dimensional dependences clear. Here $\gamma$
is a dimensionless parameter:
\begin{equation}
\gamma = \frac{\ell_0 d}{a^2}, \mbox{\ with\ } 
\ell_0 = \frac{4\pi e^2 n_b}{\varepsilon m \omega_c^2} .
\label{gamma}
\end{equation}
This introduces a new length scale $\ell_0$ which we will discuss shortly.

The mode equation Eq.~(\ref{modede}) has an analytic solution. We will
consider only long-wavelength modes, and hence drop the $k^2$ term. The
solution of Eq.~(\ref{modede}), given $\omega$, is
\label{fs}
\begin{equation}
f(x) = \left\{ \begin{array}{ll}
A e^{x/\lambda} , & x<0 \\
J_0\left(\xi\sqrt{1-x/a}\right), & 0<x<a,\end{array}\right.
\end{equation}
where
\begin{eqnarray}
\lambda &=& a \sqrt{
  \displaystyle\frac{\gamma\omega_c^2(\omega+i/\tau)}
    {\omega[\omega_c^2 - (\omega+i/\tau)^2]}}, \nonumber\\
\xi &=& 2\sqrt{\displaystyle \frac{ka\omega_c -
  \frac{\omega}{\gamma\omega_c^2} 
  [ \omega_c^2 - (\omega+i/\tau)^2] }
 {\omega+i/\tau} }.
\label{xi}
\end{eqnarray}
Here $A$ is an unknown constant. It and $\omega$ are fixed by
matching $f$ and $f'$ at $x=0$. The result is the secular equation
\begin{equation}
\frac{J_0(\xi)}{J_1(\xi)} =
 \sqrt{ \displaystyle\frac{ka\gamma\omega_c^3/\omega}
                               {\omega_c^2-(\omega+i/\tau)^2}-1}.
\label{secular}
\end{equation}
For each wave vector $k$ this has infinitely many solutions $\omega=\omega_n(k)$
giving the various edge magnetoplasmon modes\cite{AG}.
Notice that the exact eigenfrequencies have the symmetry
$\omega_n(-k)=-\omega_n^*(k)$. Eq.~(\ref{secular}) is consistent with
this symmetry. Under $k\rightarrow-k$, $\xi\rightarrow\xi^{*}$ and 
both sides of Eq.~(\ref{secular}) are complex conjugated.

We can obtain approximate solutions of the secular equation in the
limiting cases $\gamma\ll1$ and $\gamma\gg1$. These correspond to,
respectively, the limit of a very wide edge ($a\gg\sqrt{d\ell_0}$)
and a narrow edge ($a\ll\sqrt{d\ell_0}$). Here we are comparing
the edge width to a length given by the geometric mean of the
gate spacing $d$ and the new length scale $\ell_0$ in
Eq.~(\ref{gamma}). (This length scale $\ell_0$ also arose in
Volkov and Mikhailov's Wiener-Hopf treatment of semi-infinite systems
with abrupt boundaries\cite{VM}).  For a wide edge $a$ turns out to be
much greater than the bulk penetration depth $\lambda$ and the edge mode
resides primarily in the region $x>0$ where the equilibrium electronic
density is nonuniform (see Fig.~\ref{geometry}). Conversely for a
narrow edge $a$ is much less than the penetration depth and the
`edge mode' resides primarily in the bulk (where it vanishes
exponentially).  

Using parameters for GaAs we can write $\ell_0$ from Eq.~(\ref{gamma})
in the equivalent forms
\begin{equation}
\ell_0 = \frac{\nu_b\ell^2}{10\,\mbox{nm}} = \frac{66\nu_b}{\overline{B}}\,\mbox{nm},
\label{ell0}
\end{equation}
where $\ell$ is the magnetic length, $\nu_b$ is the bulk filling factor,
and $\overline{B}$ is the magnetic field in Tesla.
A system with bulk density $n_b=10^{11} \,\mbox{cm}^{-2}$
achieves $\nu_b=1$ at $B=4.13\,\mbox{T}$, so that 
$\ell_0\approx 16\,\mbox{nm}$. In Ref.~\onlinecite{e2}
$d\approx 130\,\mbox{nm}$,
so $\sqrt{d\ell_0}\approx 50\,\mbox{nm}$. This is far smaller than the
edge width $a\approx 500 \,\mbox{nm}$, and so these experiments fall
within the limiting case of a very wide edge ($\gamma\ll1$).

\subsubsection{Narrow edge}
\label{subsubsec:narrow}
First we consider the narow edge, for which
$a \ll \sqrt{d\ell_0}$ and $\gamma\gg1$.
In this limit the edge mode lies largely in the bulk into which it
exponentially decays. 

When $\gamma\gg1$ it turns out that the secular equation
Eq.~(\ref{secular}) is solved by $\xi\ll1$. 
Replace the left-hand 
of Eq.~(\ref{secular}) by the dominant term in the series expansion,
$J_0(\xi)/J_1(\xi)\approx 2/\xi$. For large $\gamma$ the result
is solved approximately by
\begin{equation}
\frac{\omega(k)}{\omega_c} = \sqrt{ \frac{\ell_0 d k^2}{1+1/(\omega_c\tau)^2}
 - \frac{1}{(2\omega_c\tau)^2}} - \frac{i}{2\omega_c\tau}
\label{omega2}
\end{equation}
for $k>0$. [For $k<0$ use $\omega(-k)= -\omega(k)^{*}$.] 
Plugging this back into Eq.~(\ref{gamma}) yields $|\xi| \sim \gamma^{-1/4}$;
this is small for large $\gamma$, consistent with our assertion above.
The bulk penetration depth becomes approximately 
$|\lambda|\sim a\sqrt{\gamma}=\sqrt{d\ell_0}$,
which here is much greater than the edge width $a$.

\begin{figure}
\begin{center}
\includegraphics*[scale=0.35]{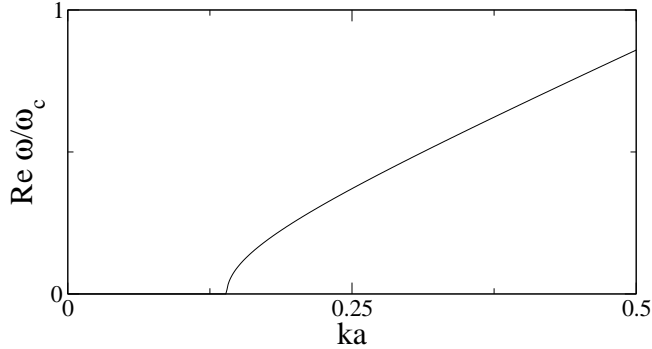}
\end{center}
\caption{Model dispersion relation with a narrow edge. Here
$1/\omega_c\tau=0.5$ , $\gamma=4$. Note that the solution is
valid only for $|\omega/\omega_c|\ll1$.}
\label{modeldispersion}
\end{figure}

Eq.~(\ref{omega2}) is the dispersion relation of a single, dominant,
mode in the limit of a narrow edge. 
It is pictured in Fig.~\ref{modeldispersion}.
Only one mode appears because we kept only one term in the expansion of
$J_0/J_1$.
For small dissipation, the mode is underdamped for
$k>1/2\omega_c\tau\sqrt{\ell_0d}$ and diffusive for smaller $k$.
In the $k \to 0$ limit, we have $\omega (k) = - i D k^2$ where the
diffusion constant is $D = s^2 \tau$ and
$s = \sqrt{\frac{4 \pi e^2 n_b d}{\varepsilon  m}}$ is the velocity of
the edge mode in the absence of dissipation (see below).
The mode found here is closely connected to the
result of Volkov and Mikhailov, who considered systems
with $a$ identically zero\cite{VM}. Modifying their
Wiener-Hopf technique to include a gate reproduces Eq.~(\ref{omega2}).

Notice an interesting aspect of the edge magnetoplasmon mode in
this narrow edge regime: its velocity is nearly independent of magnetic
field. For small dissipation, Eq.~(\ref{omega2}) becomes
\begin{equation}
\frac{\omega}{k}  \approx \omega_c \sqrt{\ell_0 d} = 
\sqrt{\frac{4\pi e^2 n_bd}{\varepsilon m}}.
\label{linearw}
\end{equation}
This behavior is quite different from that of wide edges, for which
the velocity is largely $E/B$ drift. Here the mode resides mostly in the
bulk (where it dies exponentially) and the electric field on the edge
($x>0$) plays no role.

Edge magnetoplasmons can be treated semiclassically only in
the limit $a\gg\ell$, the case  of a `smooth edge' discussed
in the introduction. The `narrow' limit $a \ll \sqrt{d\ell_0}$
described in this section thus
only gives physically reliable results when the magnetic length
$\ell$ is much smaller than the other lengths $a$, $\sqrt{d\ell_0}$.
To approach this narrow edge limit experimentally
would require a combination
of increased gate spacing $d$ and smaller electron density $n_b$ [the
latter to increase $\ell_0$; see Eq.~(\ref{gamma})].  Of course as a formal
limit of the model described by Eq.~(\ref{modede}) the limit always exists.

\subsubsection{Wide edge}
\label{subsubsec:wide}

The wide edge  has $a \gg \sqrt{d\ell_0}$ and $\gamma \ll 1$.
In this case there is little penetration into the bulk; the edge
mode lives on the outer edge $x>0$ where the equilibrium density is
inhomogeneous.

For orientation let us begin by removing the dissipation
(set $1/\tau=0$). When $\gamma\ll1$ it turns out that the
right-hand side of Eq.~(\ref{secular}) becomes small.
(We check consistency below.) Consequently
the solutions of Eq.~(\ref{secular}) are given by
\begin{equation}
\xi \approx r_n
\label{xirn}
\end{equation}
where $r_n \approx 2.405,5.520,\dots$ are the zeros of $J_0$ ($n=0,1,2,\dots$).
Using Eq.~(\ref{xi}) and (valid for long-wavelength modes) 
dropping terms second order in $\omega/\omega_c$,
this gives
\begin{equation}
\omega_n(k) = \frac{\gamma \omega_c ka}{1+\gamma r_n^2/4}.
\label{nodiss}
\end{equation}
This is an example of the general results for nondissipative systems
obtained by Aleiner and Glazman\cite{AG}. We see that that the fastest
mode ($n=1$) is nodeless. Its speed is approximately 
$\gamma\omega_ca$, which is identical with the estimate Eq.~(\ref{sest});
the speed falls as $1/B$.

Now turn on dissipation. We repeat the solution above, assuming
that the right-hand side of Eq.~(\ref{secular}) remains small; 
checking consistency at the end defines conditions (a lower bound on $ka$)
under which this procedure gives an accurate solution. 
Combining Eqs.~(\ref{xi}) and (\ref{xirn}) yields
\begin{equation}
\omega_n(k) = \gamma\omega_c\, \frac{ka-\displaystyle\frac{r_n^2}{4}\frac{i}{\omega_c\tau}}
{1+\gamma\displaystyle\frac{r_n^2}{4}-\left( \frac{\omega_n(k)}{\omega_c}+\frac{i}{\omega_c\tau}\right)^2
}.
\end{equation}
For small $\gamma$ this can be solved for $\omega_n(k)$ by iterative substitution
into the denominator. For our purposes it is sufficient to look
at the lowest order approximation,
\begin{equation}
\omega_n(k) = \gamma\omega_c \,\frac{ka-\displaystyle\frac{r_n^2}{4}\frac{i}{\omega_c\tau}}
{1+\gamma\displaystyle\frac{r_n^2}{4} + \frac{1}{\omega_c^2\tau^2} }.
\label{omega1}
\end{equation}
Once again we see the modes move slower as the number $n$ of nodes increases.
It is also now evident that increasing $n$ leads to increasing damping
($|\mbox{Im}\,\omega_n|$ increases). Thus within this model the nodeless
mode ($n=0$) is both the fastest and least damped, and will dominate
time-of-flight experiments such as that of Ernst \textit{et al.}

It remains to check consistency. Using Eq.~(\ref{omega1}), we find that
the right-hand side of Eq.~(\ref{secular}) is small as long as
$ka \gg r_n^2/\omega_c\tau$. This regime exists for long-wavelength modes
as long as the dissipation is sufficiently small. Then Eq.~(\ref{omega1})
is the dispersion relation for long wavelength modes above a lower
bound ($ka \gg r_n^2/\omega_c\tau$).  For such modes
the bulk penetration depth $|\lambda|\approx a/(\omega_c\tau ka)^{1/2}$
is much less than $a$, as claimed; thus the mode resides largely on the
outer edge.

The approximations leading to Eq.~(\ref{omega1}) fail for extremely
long wavelengths ($ka \lesssim 1/\omega_c\tau$). Consider the limit
$ka \ll 1/\omega_c\tau$.  From Eq.~(\ref{xi}) $\xi\ll1$ and
so $J_0(\xi)/J_1(\xi)\approx2/\xi$.  Then from Eq.~(\ref{secular})
we find that at very long wavelenths the mode becomes diffusive:
$\omega = -iDk^2$ with $D=\tau d 4\pi e^2 n_b/\varepsilon m$. 
Exactly the same diffusive limit occurs for small $k$ in the case
of a narrow edge, from Eq.~(\ref{omega2}).

\subsection{Oscillations in the quantum Hall regime}
\label{subsec:oscillations}

Our last step is to model the variation of the dissipation with
magnetic field and to show how this variation influences the edge
magnetoplasmons.  As disucssed in Section II, we treat $1/\tau$
as a global property of the edge which depends only on the bulk
filling factor $\nu_b$. 
This incorporates one crucial aspect of quantum mechanics
(the loss of dissipation on a Hall plateau) into an otherwise classical
treatment of edge magnetoplasmons. We expect this to at least qualitatively
capture the role of dissipation for the edge magnetoplasmons. It turns out
that the effect is appreciable.

\begin{figure}
\begin{center}
\includegraphics*[scale=0.35]{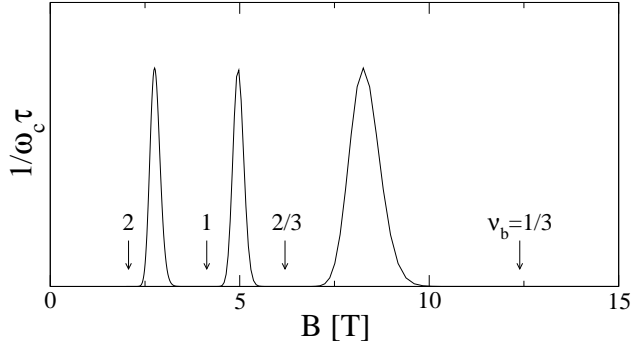}
\end{center}
\caption{Model for $1/\omega_c\tau$. The peak value between Hall plateaus is
$(1/\omega_c\tau)_{\mathrm{max}}$.}
\label{gaussian}
\end{figure}

We model the magnetic field dependence of $1/\tau$ in the following manner:
\begin{equation}
\frac{1}{\omega_c\tau} = \left(\frac{1}{\omega_c\tau}\right)_{\mathrm{max}}
g(\nu_b) .
\label{diss}
\end{equation}
Here $g(\nu_b)$ is a function which is zero on Hall plateaus and rises to a
maximum value of 1 between plateaus. In such a phenomenological model
one can include as many Hall plateaus as desired. In the following we
keep Hall plateaus at $\nu_b=2,1,2/3,1/3$. The detailed shape of $g(\nu_b)$
is unimportant. For convenience we used a Gaussian, pictured in 
Fig.~\ref{gaussian}. At typical experimental temperatures the constant
$(1/\omega_c\tau)_{\mathrm{max}}$ should be of order 1.

Switching on the dissipation $1/\omega_c\tau$ between Hall plateaus
both damps the edge magnetoplasmons and changes their speed. 
For example, for a wide edge consider the model dispersion relation in
Eq.~(\ref{omega1}). Between Hall plateaus $\mathrm{Im}\,\omega_n(k)$ 
becomes nonzero and the mode is damped. At the same time
$\mathrm{Re}\,\omega_n(k)$ gets smaller --- the mode slows down as
it is damped. Thus switching on and off the dissipation as the system 
moves off and on Hall plateaus causes simultaneous oscillations in
the mode speed and strength. The strength of the oscillations grows with
$(1/\omega_c\tau)_\mathrm{max}$ and can be appreciable (see the next
section). 

The oscillations could be quite large for a narrow edge. In a
time-of-flight experiment an initial voltage pulse triggers an electron
density pulse which travels along the edge. Suppose the density pulse
is a wave packet centered around a characteristic
wave vector $k_0$. For small dissipation, modes near $k_0$ will be 
underdamped and move with a speed of order $\omega_c\sqrt{\ell_0d}$
[see Eqs.~(\ref{omega2}, \ref{linearw})]. As the dissipation grows the
modes near $k_0$ can become diffusive, yielding large drops in signal
strength and velocity.

In the next section we examine the role of dissipation on the
edge magnetoplasmons with fewer approximations, continuing to
use the phenomenological model for dissipation [Eq.~(\ref{diss})],
but solving the mode equations [Eqs.~(\ref{modeeqn})] numerically.

\subsection{Dissipation and edge modes: numerical results}
\label{subsec:numerics}

We conclude by studying the effect of dissipation on the edge magnetoplasmons
by direct numerical solution of Eqs.~(\ref{modeeqn}). The simplest way to do
this is to discretize space, which converts the mode equations into an
eigenvalue problem. Since the resulting matrix to be diagonalized is complex,
some care needs to be taken in identifying the proper modes. We were guided
by the analytical results obtained above.

In this section we use parameters similar to the experiments of
Ref.~\onlinecite{e2}. Here the bulk electronic density 
$n_b=10^{11}\,\mathrm{cm}^{-2}$, the gate spacing $d=130\,\mathrm{nm}$,
edge width $a=1.2\,\mu\mathrm{m}$, and for GaAs $m=0.067\,\mathrm{m}_e$,
$\varepsilon=12.4$. As mentioned above, these parameters
put us well within the limit of wide edges. The modes are
not very sensitive to the details of the edge confinement. Here we use a
parabolic confinement; this leads to an edge density
profile $n_0(x)=n_b(1-x^2/a^2)$ for $0<x<a$. Because
$|\omega|\ll\omega_c$ for the fields considered it was sufficient to
take the DC limit of the conductivities Eq.~(\ref{sigma}).
The mode equations are solved for the eigenvalues $\omega$ and corresponding
mode densities $f(x)$ as a function of $B$ and $k$.

\begin{figure}
\begin{center}
\includegraphics*[scale=0.35]{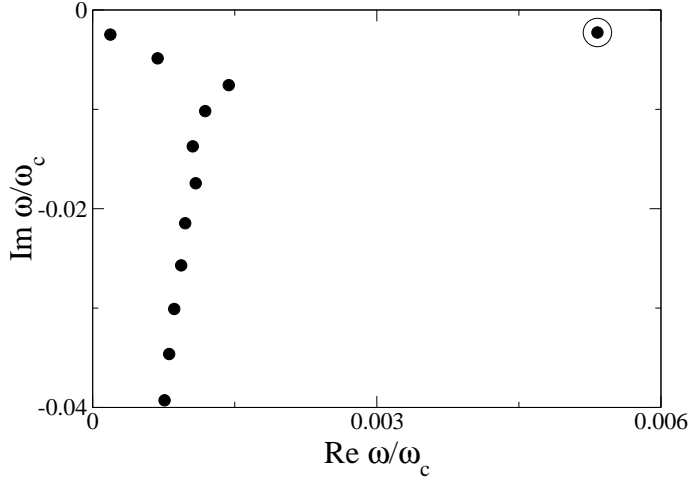}
\end{center}
\caption{Eigenvalues in the presence of dissipation. Pictured are the
12 eigenvalues with the smallest $|\mathrm{Im}\,\omega|$. Notice that
one mode (circled) is both the fastest and least damped. This is the
dominant mode for time-of-flight experiments. Here
$\nu_b=1.5$, $(1/\omega_c\tau)_{\mathrm{max}}=0.3$, $k=1/a$.}
\label{eigenvalues=1_1.5}
\end{figure}

\begin{figure}
\begin{center}
\includegraphics*[scale=0.35]{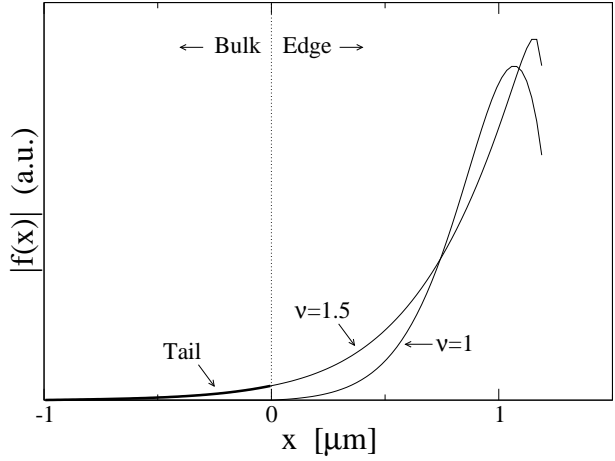}
\end{center}
\caption{Mode densities for $\nu_b=1$ and $\nu_b=1.5$. In the absence
of dissipation ($\nu_b=1$) the mode resides entirely on
the edge. In the presence of dissipation ($\nu_b=1.5$) the mode has
an exponentially decreasing tail in the bulk. 
Here $(1/\omega_c\tau)_{\mathrm{max}}=0.05$, $k=1/a$.}
\label{modes_nu=1_1.5}
\end{figure}

Diagonalization of the discretized form of Eqs.~(\ref{modeeqn}) yields
as many modes as there are grid points (typically 200-500).
In the presence of dissipation
the eigenvalues become complex. An example is pictured in 
Fig.~\ref{eigenvalues=1_1.5}, which shows the 12 eigenvalues
with the smallest dissipation for a particular run. Notice that
one mode in Fig.~\ref{eigenvalues=1_1.5}
is singled out as both the fastest and least
damped. This mode dominates time-of-flight experiments.
As the dissipation increases this mode continues to be identifiable;
the fastest mode continues to have relatively small damping. In the following we
focus on this dominant mode.
In the absence of dissipation the dominant mode resides entirely on the
edge. Switching on dissipation causes it to penetrate
into the bulk, where it dies exponentially (Fig.~\ref{modes_nu=1_1.5}).

\begin{figure}
\begin{center}
\includegraphics*[scale=0.35]{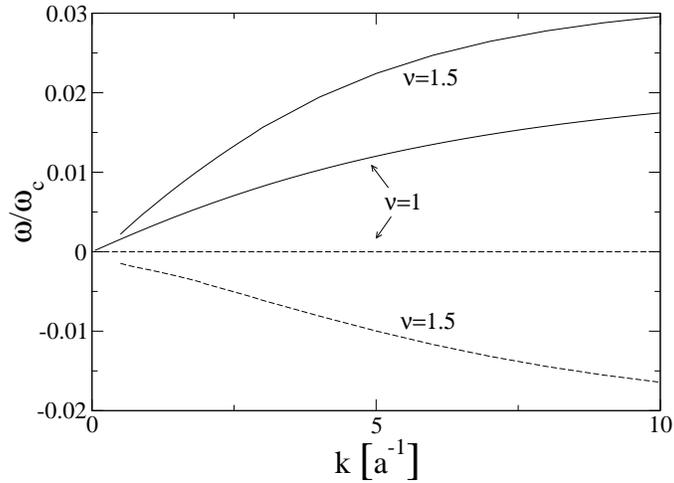}
\end{center}
\caption{Dispersion relation $\omega(k)$ for the dominant mode for
a nondissipative case ($\nu=1$) and a dissipative case ($\nu=1.5$).
The solid lines are $\mathrm{Re}\,\omega$ and the broken lines are
$\mathrm{Im}\,\omega$.
For very small $k$ in the presence of dissipation the mode becomes
diffusive and difficult to identify within the numerical results.
Here $(1/\omega_c\tau)_{\mathrm{max}}=0.3$.}
\label{eigenvalues_nu=1_1.5}
\end{figure}

Sweeping the wave vector $k$ yields a dispersion relation $\omega(k)$
for the dominant mode. This is pictured for two cases (no dissipation
at $\nu_b=1$, dissipation at $\nu_b=1.5$) in Fig.~\ref{eigenvalues_nu=1_1.5}.

\begin{figure}
\begin{center}
\includegraphics*[scale=0.35]{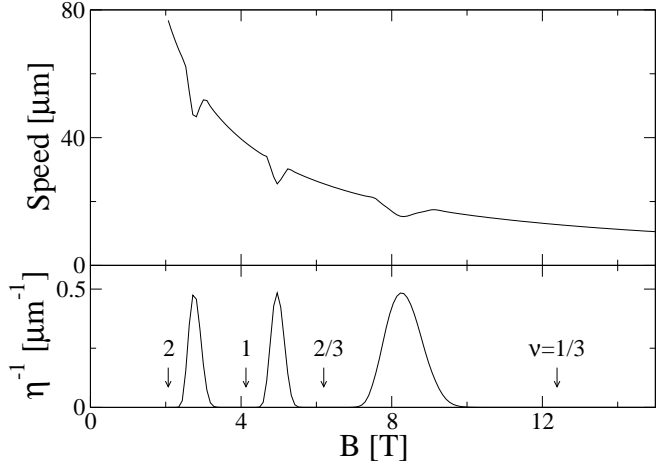}
\end{center}
\caption{Speed and damping of the fastest mode as a function of
magnetic field. Increasing dissipation between Hall plateaus leads
to more damping (shown in the lower panel as an inverse decay length)
and oscillations in speed (shown in the upper panel).
Here $(1/\omega_c\tau)_{\mathrm{max}}=1$.}
\label{speeds}
\end{figure}

The speed and damping of the dominant
mode as a function of magnetic field are shown in Fig.~\ref{speeds}.
In a time-of-flight experiment wave packets centered at some $k_0$
travel along the edge. The value of $k_0$ is set by geometrical
factors (such as the probe dimensions) and the time dependence of
the voltage pulse. Typically a voltage probe is about a micron wide
and is also about a micron from the edge of the electronic system.
This is approximately the size of the edge width $a$. Consequently
we assume that $k_0$ is of order $1/a$. (The results below do not
change qualitatively as long as $k_0$ is in the linear portion of
the dispersion relation for small dissipation.)
The speed shown in the upper panel of Fig.~\ref{speeds} is the slope
\begin{equation}
s = \left.\frac{d\omega}{dk}\right|_{k=k_0}~,
\end{equation}
where $k_0=1/a$. Superimposed on the $1/B$ trend in velocity
are oscillations caused by switching on dissipation between plateaus.
We report the mode's strength using an inverse decay length
\begin{equation}
\eta^{-1} = |\mathrm{Im}\,\omega(k_0)| / s .
\end{equation}
In the absence of dissipation $\eta^{-1}=0$. Increasing dissipation
leads to increasing $\eta^{-1}$, as shown in the lower panel of 
Fig.~\ref{speeds}. 

The oscillations in speed and damping evident in Fig.~\ref{speeds} are
large enough to be experimentally visible. They are at least qualitatively
similar to the oscillations in signal strength and mode speed observed in
the experiments of Ernst \textit{et al.},\cite{e2}
and may explain their results.

\section{Summary}

We have presented a semiclassical treatment of edge magnetoplasmons
in the quantum Hall regime, in the presence of dissipation. Our
treatment of the edge magnetoplasmons is essentially classical,
with dynamics arising from the continuity equation and self-consistent
electric fields. Quantum mechanics enters the calculation in two places. First,
energy gaps associated with the quantum Hall effect lead to incompressible
stripes within the largely incompressible edge. We find these stripes to have
very little effect on the edge modes. But the physics of the quantum Hall
effect also enters in a second place, as a
dissipative contribution to the conductivity tensor. 
The dissipation influences the modes in a natural way.
Between Hall plateaus, switching on dissipation leads to increased
damping and a diminished mode velocity. This behavior is
similar to the expected role of dissipation in other signatures of electronic
plasma oscillations, such as Surface Acoustic Wave (SAW) experiments.
The interesting aspect of quantum Hall effect systems is that the
dissipation can be turned on and off by varying the magnetic field.

\begin{acknowledgments}
We are thankful for useful conversations with Gabriele Ernst,
Klaus von Klitzing, and Olle Heinonen.
This work was supported by the NSF under Grants No. DMR-9972683 and DMR-0074959.
\end{acknowledgments}

\end{document}